\documentclass[lettersize,journal]{IEEEtran}
\usepackage{amsmath,amssymb,amsfonts}
\usepackage{algorithmic}
\usepackage{algorithm}
\usepackage{array}
\usepackage[caption=false,font=normalsize,labelfont=sf,textfont=sf]{subfig}
\usepackage{textcomp}
\usepackage{stfloats}
\usepackage{url}
\usepackage{verbatim}
\usepackage{graphicx}
\usepackage{cite}

\usepackage{widetext}

\begin{document}

\title{Time-domain Response of Supercapacitors using their Impedance Parameters and Fourier Series Decomposition of the Excitation Signal}

\author{
\IEEEauthorblockN{Anis Allagui$^{1,2}$, Mohammed E. Fouda$^3$, \IEEEmembership{Senior Member IEEE}, Ahmed Elwakil$^{3,4,5}$, \IEEEmembership{Senior Member IEEE} and Costas Psychalinos{$^6$}, \IEEEmembership{Senior Member IEEE}}

\thanks{{$^1$Dept. of Sustainable and Renewable Energy Engineering, University of Sharjah, United Arab Emirates (aallagui@sharjah.ac.ae)} \\ 
{$^2$Dept. of Mechanical and Materials Engineering, Florida International University, Miami, FL33174, United States (aallagui@fiu.edu)}\\
{$^3$Nanoelectronics Integrated Systems Center (NISC), Nile University, Giza, Egypt}\\
{$^4$Dept. of Electrical Engineering, University of Sharjah,  United Arab Emirates (elwakil@ieee.org)}\\
{$^5$Dept. of Electrical Engineering, University of Calgary, Alberta, Canada (ahmed.elwakil@ucalgary.ca)}\\
{$^6$Dep. of Physics, University of Patras, Patras, Greece (cpsychal@upatras.gr)}
}
}



\maketitle

\begin{abstract}
Supercapacitors are mostly recognized for their high power density capabilities and fast response time when compared to secondary batteries. However, computing their power in response to a given excitation using the standard formul\ae\;of capacitors is misleading and erroneous because supercapacitors are actually non-ideal capacitive devices that cannot be characterized with a single constant capacitance. In this study we show how to estimate accurately the time-domain power and energy of supercapacitors in response to any excitation signal represented in terms of its Fourier series coefficients  with the sole knowledge of the frequency-domain impedance parameters of device. The presented theory is first verified and validated with simulations conducted on an equivalent fifth-order $RC$ circuit emulating the behavior of a fractional circuit consisting   of a resistance ($R_{s}$) in series with a  constant phase element (CPE) of fractional impedance $Z_{\text{CPE}}={1}/{C_{\alpha}s^{\alpha}}$. Then we do the same for a commercial supercapacitor modeled as an $R_s$-CPE circuit, and subjected to both a periodic triangular voltage waveform and a random voltage excitation.  The results are conclusive and very promising for adopting the proposed procedure to estimate the power and energy performance of supercapacitors in response to real-world charging and discharging signals.  
\end{abstract}


\section{Introduction}
\IEEEPARstart{T}{he}  main features of supercapacitors or electrochemical capacitors as electrical energy storage devices are the fact that they are practically maintenance-free with high degree of reversibility, extended range of operation temperature, and lifetimes in the order of millions of charge/discharge cycles \cite{shao2018design}. In terms of energy storage capabilities, supercapacitors are still behind when compared to state-of-the-art secondary batteries but the gap is getting narrower  with more recent research and development \cite{zhang2022high, shao2018design, yan2014recent, https://doi.org/10.1002/admt.202200459}.   
However, frequency-domain data of most supercapacitors, obtained by electrochemical impedance spectroscopy (EIS) characterization,    point out to a non-ideal capacitive behavior \cite{kim2011advanced, vivier2022impedance}.  The frequency dispersion of their electrical parameters has been attributed to different physical reasons, including surface effects (roughness, heterogeneity), porous/fractal structure of constituting electrodes, and slow ionic diffusion  within the pores of the electrodes \cite{Pajkossy}. 
Nonetheless, this frequency-dependent performance of supercapacitors affecting their  operating characteristics at high rates does not prevent them from being used in numerous  applications including  backup power, pulse power and hybrid power systems\;\cite{conway2002power, chang2016active, zhang2021multi}. They can be used either as the sole energy storage reservoir depending on the system requirements, or in combination with secondary batteries or fuel cells to optimize cost, life time and run time in electric vehicles applications for example\;\cite{PE2,conway2002power, dasari2020simple}. 

 To account for the non-ideal  behavior of supercapacitors, the  constant phase element (CPE) is usually invoked in modeling their spectral response   \cite{noori2019towards, EES, nanoE, 10.1149/1945-7111/ac621e, zhang2022fractional}. The CPE is a mono-order fractional capacitor of  impedance defined as 
\begin{equation}
Z_{\text{CPE}}=\frac{V(s)}{I(s)}=\frac{1}{C_{\alpha}s^{\alpha}}
\label{eqZPE}
\end{equation} 
where $s^{\alpha}=\omega^{\alpha} \angle\,\alpha\pi/2$,  $0<\alpha \leqslant 1$ and $C_{\alpha}$ is a pseudocapacitive parameter in units of F\,s$^{\alpha-1}$. The deviation of the  dispersion coefficient $\alpha$ from 1.0 is a measure of the departure from ideal capacitance behavior.  However, plausible physical interpretation of the CPE is  still obscure and perplexing, even after decades of research and discussions \cite{lasia2022origin,sadkowski2000ideal, ieee, ieeeted2}. 
Additionally, analyzing the response of these devices (and other similar ones involving CPEs) from the time domain perspective  can be quite challenging.   Closed-form solutions can be obtained only in a limited number of cases such as when the excitation signal is a constant voltage/current \cite{energy2015,allagui2021possibility}, linear ramp voltage/current \cite{IM,energy2015,EC2015} or power law functions \cite{allagui2021possibility}. This leaves the door open for analyzing the behavior of such devices when different forms of excitations are applied, as it can be frequently encountered in real world applications. 
 In particular, given that supercapacitors are mainly used for their high power density capabilities for load leveling or supplying high energy pulses for short durations,  it is important to have the correct tools to compute such power in response to any type or form of charging or discharging excitation. 
 
  In this work, we demonstrate a  systematic method to  compute time-domain supercapacitor response with knowledge of (i) its frequency-domain impedance model parameters, and (ii) the Fourier series decomposition of the excitation signal that can be of any arbitrary form. 
  For doing so, i.e. the transition from frequency-domain to time-domain, it is required to invoke the (discrete) convolution theorem   \cite{ieeeted}. 
 The advantage of this approach is first its   applicability to any input excitation. In addition, this method  is valid   for any  frequency-domain model of the supercapacitor, unlike closed-form expressions which are derived for a specific model only and for specific excitations.  Power and energy calculations on an $R_s$-CPE circuit in response to a full-wave rectified voltage signal is provided as an example,  and validated using a fifth-order distributed $RC$ circuit emulator. Experimental results on a commercial supercapacitor in response to a periodic triangular voltage input and a random voltage excitation (i.e. random amplitudes and random durations for each step) are presented and discussed in conjunction with the proposed computational procedure.

\section{Theory\label{sec:proposed concept}}

\begin{figure*}[t]
\begin{center}
\includegraphics[angle=-90,width=.9\textwidth]{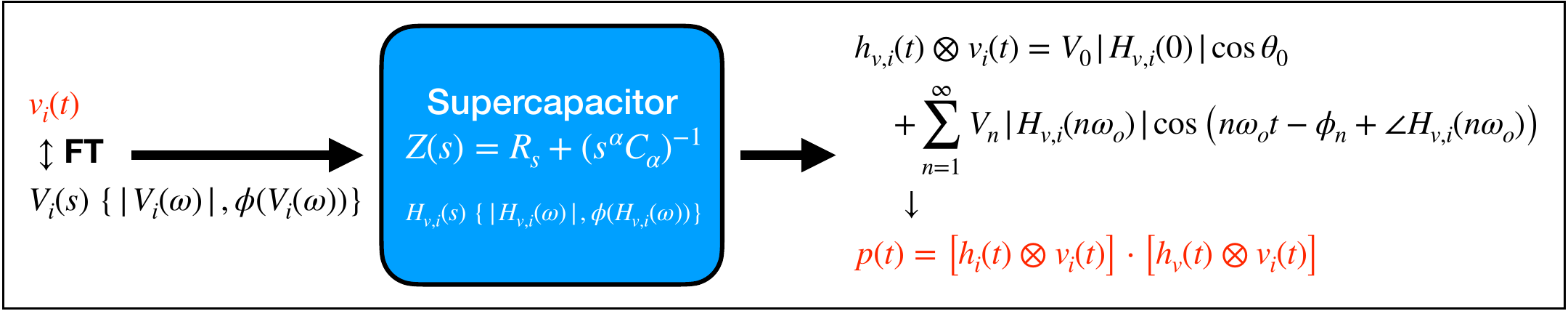}
\caption{Schematic diagram of the steps to follow for computing the instantaneous power, $p_c(t)$ on the CPE (Eq.\;\ref{eq:p}) in a supercapacitor modeled as an $R_s$-CPE in response to a random voltage excitation, $v_i(t)$ (Eq.\;\ref{eq:vit})}
\label{fig1}
\end{center}
\end{figure*}

Consider a linear, time-invariant (LTI) capacitive device modeled, for example, by an  equivalent circuit composed of  resistor $R_{s}$ in series with a CPE. The set of  parameters $R_s$, $C_{\alpha}$ and $\alpha$ are known a priori,  and can be obtained by complex nonlinear fitting of Eq.\;\ref{eqZPE} to the measured spectral impedance data.  When a voltage excitation  $V_i(s)$ defined in the frequency domain is applied across this $R_s$-CPE model,  it can be shown that the voltage across the CPE alone $V_{c} (s)$
is given by \cite{conv}:
\begin{equation}
V_{c}(s)=\left(\frac{1}{1+R_{s}C_{\alpha}s^{\alpha}}\right)V_{i}(s)=H_{v}(s) \cdot V_{i}(s)\label{eq:Hi}
\end{equation}
The multiplication operation in the frequency-domain translates  
to a convolution operation in the time domain; i.e.
\begin{equation}
v_{c}(t)
= \int_{0}^{\infty} h_v(\tau)  v_i(t-\tau) d\tau  
=h_{v}(t)\otimes v_{i}(t)
\label{eqvc}
\end{equation}
where lower case and higher case letters are used to denote functions that are Laplace transform pairs, i.e. $F(s) = \mathcal{L}\{f(t)\} = \int_{0}^{\infty} f(t) e^{-st } dt$. 
The  impulse response function $h_{v}(t)$ 
is obtained by inverse Laplace transform, and  is given by  \cite{mathai2008special}:
\begin{equation}
h_{v}(t)=\frac{t^{\alpha-1}}{R_{s}C_{\alpha}}E_{\alpha,\alpha}\left(-\frac{t^{\alpha}}{R_{s}C_{\alpha}}\right)
\label{eq4}
\end{equation}
where  $E_{\alpha,\alpha}\left(z\right)$
is the  two-parameter Mittag-Leffler
function:
\begin{equation}
{E}_{\alpha,\beta} ( z ) := \sum\limits_{k=0}^{\infty} \frac{z^k}{\Gamma(\alpha k + \beta)}  \quad (\alpha,\beta \in \mathbb{C}, \mathrm{Re}({\alpha, \beta})>0)
\label{eqML}
\end{equation}
Similarly,  the  frequency-domain current passing through the supercapacitor
in this case is given by:
\begin{equation}
I_{c}(s)=\left(\frac{C_{\alpha}s^{\alpha}}{1+R_{s}C_{\alpha}s^{\alpha}}\right)V_{i}(s)=H_{i}(s) \cdot V_{i}(s)\label{eq:Hv}
\end{equation}
which, in the time-domain corresponds to:
\begin{equation}
i_{c}(t)=h_{i}(t)\otimes v_{i}(t)
\label{eqic}
\end{equation}
and
\begin{equation}
h_{i}(t)=\frac{1}{R_{s}}E_{\alpha}\left(-\frac{t^{\alpha}}{R_{s}C_{\alpha}}\right)
\label{eq8}
\end{equation}
where $E_{\alpha}\left(-x\right)=E_{\alpha,1}\left(-x\right)$ 
 is  the single-parameter Mittag-Leffler
function.

To find the energy stored in the supercapacitor device, we need first to  calculate
the instantaneous power $p_c(t)=i_{c}(t)v_{c}(t)$ on the CPE   as:
\begin{equation}
p_c(t)=\left[h_{i}(t)\otimes v_{i}(t)\right]\cdot\left[h_{v}(t)\otimes v_{i}(t)\right]\label{eq:p}
\end{equation}
On the other hand, a voltage excitation $v_{i}(t)$ can be represented by a Fourier series as:  
\begin{equation}
v_{i}(t)=V_{0}+\sum_{n=1}^{\infty}V_{n}\cos\left(n\omega_{o}t + \phi_{n}\right)
\label{eq:vit}
\end{equation} 
where $V_{0}$ is a dc term that may or may not be present, depending
on the type of signal, and $\omega_{o}$ is the fundamental frequency.
The result of the convolution operation of any function $h(t)$ with
this $v_{i}(t)$ is given by \cite{proakis2001digital}:
\begin{align}
h(t)\otimes &v_{i}(t)=V_{0}|H(0)|\cos\theta_{0} \nonumber+\\& \sum_{n=1}^{\infty}V_{n}|H(n\omega_{o})|\cos\left(n\omega_{o}t + \phi_{n}+\angle H(n\omega_{o})\right)\label{eq:conv}
\end{align}
where $|H(n\omega_{o})|$ is the magnitude of the transfer function
$H(s)$ and $\angle H(n\omega_{o})$ is its phase angle evaluated
at each harmonic frequency. From Eqs.\;(\ref{eq:Hi}) and (\ref{eq:Hv}), the magnitudes  of $H_{v}(s)$ and $H_{i}(s)$  are given by:
\begin{align}
&|H_{v}(n\omega_{o})|=\frac{1}{D(n\omega_{o})}, \\
& |H_{i}(n\omega_{o})|=\frac{C_{\alpha}(n\omega_{o})^{\alpha}}{D(n\omega_{o})}
\end{align}
where 
\begin{equation}
 D(n\omega_{o})  =  \sqrt{1 + 2(n\omega_{o})^{\alpha}R_{s}C_{\alpha} \cos \left(\frac{\alpha\pi}{2}\right) + (n\omega_{o})^{2\alpha}R_{s}^{2}C_{\alpha}^{2}}\label{eq:D}
\end{equation}
and their respective phases are given by: 
\begin{align}
&\angle H_{v}(n\omega_{o})=-\tan^{-1} \left(\frac{(n\omega_{o})^{\alpha}R_{s}C_{\alpha}\sin\left(\frac{\alpha\pi}{2}\right)}{1+(n\omega_{o})^{\alpha}R_{s}C_{\alpha}\cos\left(\frac{\alpha\pi}{2}\right)}\right)\\
&\angle H_{i}(n\omega_{o})=\frac{\alpha\pi}{2}-\tan^{-1} \left(\frac{(n\omega_{o})^{\alpha}R_{s}C_{\alpha}\sin\left(\frac{\alpha\pi}{2}\right)}{1+(n\omega_{o})^{\alpha}R_{s}C_{\alpha}\cos\left(\frac{\alpha\pi}{2}\right)}\right)
\label{eq:mp}
\end{align}
A schematic diagram summarizing the proposed procedure to compute the instantaneous power in an $R_s$-CPE-equivalent model is shown in Fig.\;\ref{fig1}. 

\begin{figure*}[!t]
\begin{center}
\includegraphics[width=\textwidth]{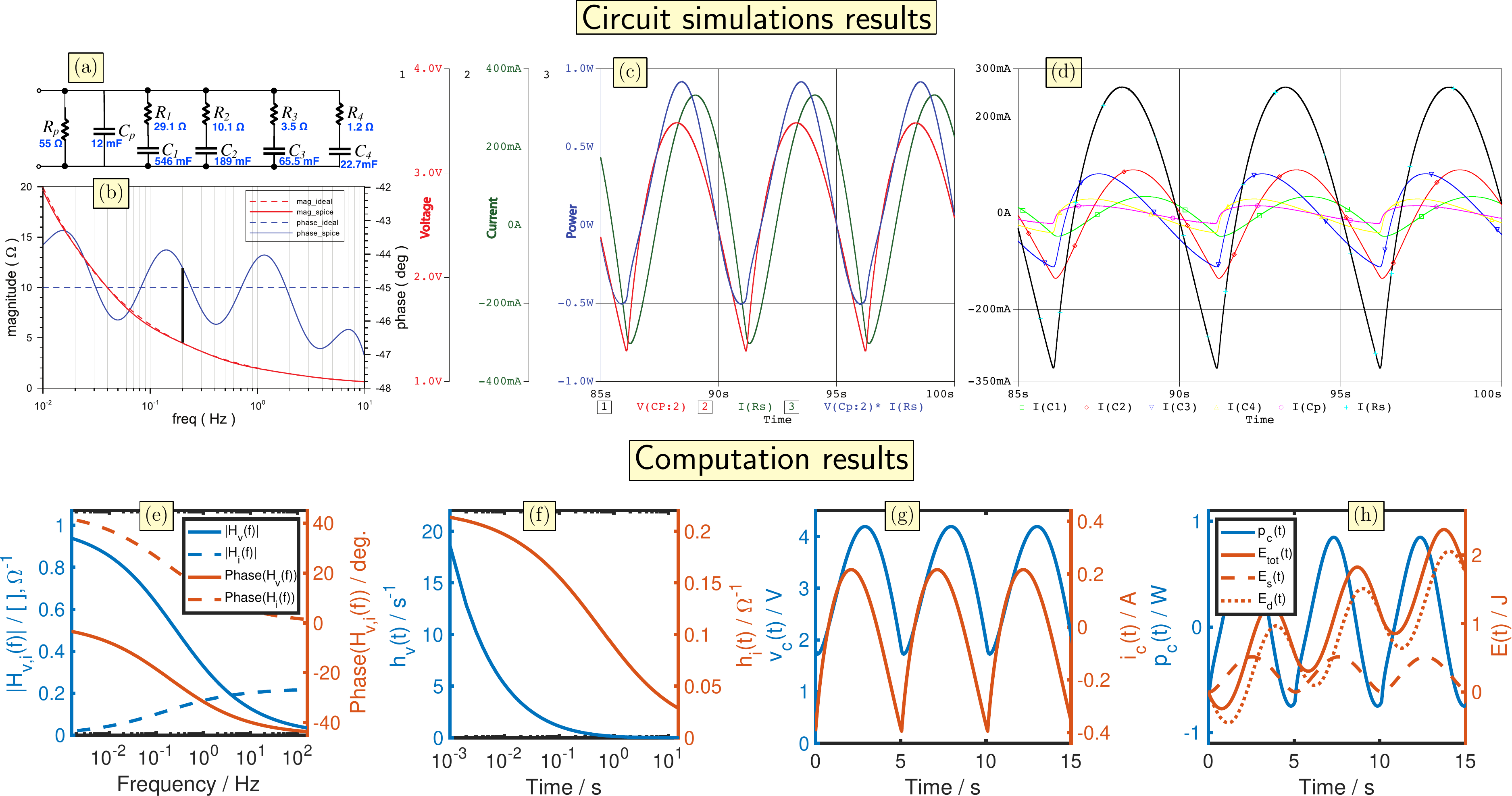}
\caption{
In 
(a) we show the integer-order $RC$ equivalent circuit of the  $R_s$-CPE model, with the corresponding component values.
(b) Simulated magnitude and phase response of the $RC$ equivalent circuit; the operating point at 200\,mHz is marked by a vertical
line. 
(c) Spice simulation of the  voltage, current, power and average energy waveforms vs. time, and in (d) we show the distribution of currents on all five capacitors of the $RC$   circuit.  
Results obtained using the $R_s$-CPE circuit  considered with parameters  
$C_{a}=\text{0.2\,F}/\sqrt{\text{sec}}$, $\alpha=$\,0.5 and 
$R_{s}=4.5\,\varOmega$ over the frequency range 20\,mHz to
 10\,Hz: 
(e) plots of   magnitudes and phase angles vs. frequency of the transfer functions $H_v(s)$ and  $H_i(s)$, and 
(f)    corresponding  step-response functions $h_v(t)$, $h_i(t)$. 
(g) Plots   of the waveforms  $v_c(t)$ and $i_c(t)$ in response to  a full-wave rectified input signal (5\,V amplitude, 1.24 rad\,s$^{-1}$ frequency), and 
(h) plots of  instantaneous  power and energy (total, stored and dissipated) on the CPE as a function of time.
}
\label{fig2}
\end{center}
\end{figure*}

It is important to emphasize again that all parameters needed to compute Eq.\;\ref{eq:conv} are frequency-domain quantities that are obtained a priori from spectral impedance fitting. 
Also, we note that while $H_{v}(s)$ represents a lowpass
filter response with cutoff frequency $\omega_{c}=(R_{s}C_{\alpha})^{-1/\alpha}$,
$H_{i}(s)$ represents a highpass filter with the same cutoff frequency
(see Fig.\;\ref{fig2}(e)). Therefore when using Eq. \ref{eq:conv}, the
upper limit of the summation can roughly be limited to $n_{max}=10\omega_{c}/\omega_{o}$
(instead of $\infty$) due to the lowpass response while the lower
limit can be limited to $n_{min}=\omega_{c}/10\omega_{o}$ due to
the highpass response. Caution however should be exercised to ensure
the validity of the $R_{s}$-CPE model  within the frequency range
$[\omega_{c}/10,10\omega_{c}]$, meaning that the values of $R_{s},C_{\alpha}$
and $\alpha$ shall remain relatively unchanged in this frequency
range. Due to the combined lowpass and highpass filter effects, the reactive power achieves its maximum value near
the cutoff frequency $\omega_{c}$, in-line with the
results reported by Fouda et al. \cite{energy2015}.

It is also worth noting that the convolution
procedure is actually independent of the circuit used to model the
supercapacitor, and is therefore more general than the approach used 
in our previous contribution \cite{linCurrent}. In particular, if the $R_{s}$-CPE model is changed
to any other model, it is required to find the new corresponding transfer functions $H_{v}(s)$
and $H_{i}(s)$, with their respective magnitudes and phases.

Having calculated the instantaneous power $p_c(t)$ given in Eq.\;\ref{eq:p}, it is now possible to obtain the accumulated energy on the CPE by the time integral $E_{\text{tot}}=\int p_c(t) dt$. The stored and dissipated energies are computed from the  Fourier series decomposition of $p_c(t)$ into sine and cosine functions, as we shall demonstrate below.

\section{Results and discussion}

\subsection{Validation via circuit simulations}

To verify and validate the procedure outlined above (Fig.\;\ref{fig1}) for computing the instantaneous power and accumulated stored energy on a capacitive  $R_s$-CPE   system from its impedance parameters,  we first generated  the response of an $RC$ transmission-line circuit emulating an $R_s$-CPE   circuit. 
A schematic of the $RC$ circuit (of  fifth order) is depicted in Fig.\;\ref{fig2}(a) along with the values of its components. It was  designed to be equivalent to an $R_s$-CPE circuit of parameters:   
$C_{a}=\text{0.2\,F}/\sqrt{\text{sec}}$, $\alpha=\text{0.5}$ and 
$R_{s}=\text{4.5}\,\varOmega$ over the frequency bandwidth 20\,mHz-10\,Hz.  
 The impedance magnitude and phase responses of the   $RC$ circuit were simulated via Spice software, and are shown in Fig. \ref{fig2}(b). 
 It can be seen that a maximum deviation in phase from    the  expected  response of 45$^{\circ}$  is less than 1.8$^{\circ}$  over the frequency
range   10\,mHz to 10\,Hz, attesting to the excellent $RC$  approximation of the fractional circuit model under test. The selected operating frequency of the $RC$ circuit  at 200\,mHz (corresponding to $\omega_{c}\approx$\,1.24\,rad\,s$^{-1}$) is indicated  in  the figure by a vertical line. 

We take a full-wave rectified  voltage as an input signal (i.e. $v_{i}(t)$ in Fig.\;\ref{fig1}) with amplitude $A=\text{5\,V}$
and frequency $\omega_{o}=$\,1.24\,rad\,s$^{-1}$. The Spice simulated voltage, current, and 
power  waveforms in response to such an excitation  are plotted in Fig.\;\ref{fig2}(c). 
 It is also interesting to observe in Fig.\;\ref{fig2}(d) the distribution of currents (representative
also of the charge distribution) on all five capacitors of the equivalent
circuit compared to the overall current in the device shown in Fig.\;\ref{fig2}(c).  It is clear that
the overall device performance (current through $R_s$ plotted in black color) is directly related to the phase shifting and amplitude scaling in the current waveforms through the five constituting capacitors of the CPE resulting
from the distributed time constants of the circuit (i.e. $\tau_i = R_iC_i$, $i=1,2,3,4,p$). 
The amplitude scaling and phase shifting
are of course frequency-dependent and therefore the operating frequency of the device is crucial for evaluating its actual performance. In the case of periodic signals, this is easy to do. However, for non-periodic signals, the energy storage
capability is predominantly controlled by the strength of the low frequency harmonics.

Now with the same $R_s$-CPE impedance model values $C_{a}=\text{0.2\,F}/\sqrt{\text{sec}}$,  $\alpha=$\,0.5 and  $R_{s}=4.5\,\varOmega$, we proceed with the computational steps summarized  in  Fig.\;\ref{fig1}.  
First,  the full-wave rectified voltage input signal ($v_{i}(t)$) can be 
described by the Fourier series: 
\begin{equation}
v_{i}(t)=\frac{2A}{\pi}-\frac{4A}{\pi}\sum_{n=1}^{\infty}\frac{\cos(n\omega_{o}t)}{4n^{2}-1}
\end{equation}
where $A$ is the amplitude of the signal and $\omega_{o}=2\pi/T$ with $T$ being the period of the signal (half the period of the non-rectified sinusoid). 
A full-wave rectified signal can be easily obtained from power-line voltages at 50/60\,Hz using a step-down transformer and a bridge rectifier. These frequencies are however too high for current supercapacitor energy storage applications, which in general do not exceed a few Hz.
 
Second, and for illustration purposes, plots of the magnitudes and phases  of both 
 voltage and current transfer functions (Eqs.\;\ref{eq:Hi} and\;\ref{eq:Hv} using $C_{a}=\text{0.2\,F}/\sqrt{\text{sec}}$,  $\alpha=$\,0.5 and  $R_{s}=4.5\,\varOmega$)  are  shown in Fig.\;\ref{fig2}(e).
 The corresponding  step-response functions $h_{v}(t)$ (Eq.\;\ref{eq4}) and $h_{i}(t)$ (Eq.\;\ref{eq8})  
 are also shown, in Fig. \ref{fig2}(f). 
 Note that if we take $\omega_{o}=\omega_{c}\approx$\,1\,rad\,s$^{-1}$, we obtain relatively simple expressions for:
 \begin{align}
H_{v}(n\omega_{o})|_{\omega_{o}=1}&=\frac{1}{D_{0.5}(n)} \angle-\tan^{-1}\left(\frac{0.71\sqrt{n}}{1+0.71\sqrt{n}}\right) \\
H_{i}(n\omega_{o})|_{\omega_{o}=1}&=\frac{1.471\sqrt{n}}{D_{0.5}(n)} \angle\frac{\pi}{4}-\tan^{-1}  \left(\frac{0.71\sqrt{n}}{1+0.71\sqrt{n}}\right)
\end{align}
where $D_{0.5}(n)=\sqrt{1+1.417\sqrt{n}+n}$.
 The voltage $v_c(t)$ (Eq.\;\ref{eqvc}) and the current $i_c(t)$  (Eq.\;\ref{eqic})  are then  given   by:
\begin{widetext}
\begin{equation}
v_{c}(t)=h_{v}(t)\otimes v_{i}(t)  
 =\frac{10}{\pi}\left(1-\sum_{n=1}^{10}\frac{2}{(4n^{2}-1)D_{0.5}(n)}\cos\left(1.24nt-\tan^{-1}\left(\frac{0.71\sqrt{n}}{1+0.71\sqrt{n}}\right)\right)\right)
 \label{eq:vt}
\end{equation}
\begin{equation}
i_{c}(t) =h_{i}(t)\otimes v_{i}(t)  
 =\frac{-20}{\pi}\left(\sum_{n=1}^{10}\frac{1.417\sqrt{n}}{(4n^{2}-1)D_{0.5}(n)}\cos\left(1.24nt+\frac{\pi}{4}-\tan^{-1}\left(\frac{0.71\sqrt{n}}{1+0.71\sqrt{n}}\right)\right)\right)
 \label{eq:it}
\end{equation}
\end{widetext}
respectively.  Plots of $v_{c}(t)$ and $i_{c}(t)$ (Eq.\,\ref{eq:conv}) with $A=\text{5\,V}$, $\omega_{o}=$\,1.24\,rad\,s$^{-1}$, and the $R_s$-CPE parameters of the model under study are given  in Fig.\;\ref{fig2}(g). The  instantaneous power $p_{c}(t)$ calculated from the product of the two variables is depicted in Fig.\;\ref{fig2}(h), and is in very good agreement with the one obtained by Spice simulation of the equivalent $RC$ circuit (see Fig.\;\ref{fig2}(c)). 
We  note that  up to this point, the computational steps depicted in Fig.\;\ref{fig1} with 100 terms for each of the summations for $v_c(t)$ and $i_c(t)$  were conducted in a matter of 20\,ms CPU time on a MacBook Pro with a   2.2 GHz 6-Core Intel Core i7 processor using  Matlab version R2019b. 
The accumulated energy ($E_{\text{tot}}=\int p_c(t) dt$) on the CPE, along with the stored energy ($E_s$) and the dissipated energy ($E_d$) are also shown in Fig.\;\ref{fig2}(h). The latter two are computed using the numerical Fourier series decomposition of $p_c(t)$ in terms of sine and cosine functions of up to five harmonics, i.e.:
\begin{equation}
p_c(t) = a_0 + \sum\limits_{n=1}^5 a_i \cos(n\omega t) + b_i \sin(n \omega t)
\label{fourierSeries}
\end{equation}
from which $E_s(t)$ is obtained from the time integral:
\begin{equation}
E_s(t) = \int\limits_0^t dt \sum\limits_{n=1}^5 b_i \sin(n \omega t)
\label{eqEs}
\end{equation}
and $E_d(t)$ from:
\begin{equation}
E_d(t) = \int\limits_0^t dt \left[a_0+ \sum\limits_{n=1}^5 a_i \cos(n \omega t) \right]
\label{eqEd}
\end{equation}
 The coefficient of determination R-squared was 0.9995, and the root mean squared error (RMSE) was 0.0124 for the approximation of Eq.\;\ref{fourierSeries} with with 95\% confidence bounds.   
The dissipated energy across the distributed resistive elements keeps increasing with time due to the dc power term ($a_0=\text{0.108}$\,W in Eq.\;\ref{fourierSeries}, which is practically equal to $p_c(t=0)=0.106$\,W in Fig.\;\ref{fig2}(g)), whereas the stored energy in the distributed capacitive elements is centered around an average of 0.282\,J. We note that this value of $E_s$ is less than the dc stored energy $\sum_i^5 C_i v_{c_i}^2/2={3.840}$\,J  on the five capacitors of the equivalent $RC$ circuit (Fig.\;\ref{fig2}(a)) given that this circuit is designed to simulate an $R_s$-CPE (with the parameters 
$C_{a}=\text{0.2\,F}/\sqrt{\text{sec}}$, $\alpha=\text{0.5}$ and 
$R_{s}=\text{4.5}\,\varOmega$ over the frequency bandwidth 20\,mHz-10\,Hz), whereas the formula $\sum_i^5 C_i v_{c_i}^2/2$ is valid for ideal capacitors operating in dc mode. In other words, near to dc frequencies the $RC$ circuit with the given $R$ and $C$ values becomes invalid in emulating the target $R_s$-CPE system. 

In summary, this example showed clearly how the proposed method for computing time-domain electrical responses of a fractional-order capacitive system out of  (i) its frequency-domain impedance parameters, and  (ii) magnitude and phase of a periodic full-wave rectified excitation signal (Fig.\;\ref{fig1}) compare very well with simulated $RC$ equivalent circuit results. In the next example, we do the same for   commercial supercapacitor device in response to both  a periodic triangular voltage excitation and a random voltage excitation.

\subsection{Experimental results on a commercial supercapacitor}

\begin{figure}[!t]
\begin{center}
\includegraphics[]{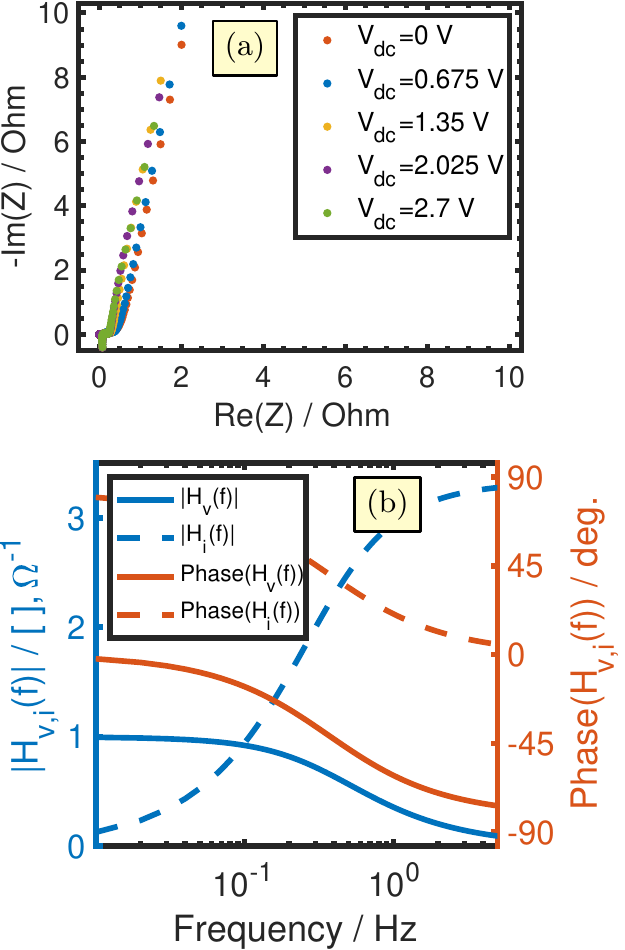}
\caption{
(a) Nyquist impedance plots of real vs. imaginary parts of the PowerStor supercapacitor at different dc voltage biases; fitted to an $R_s$-CPE model from 3.4\,Hz down to 10\,mHz gave the average values of  $R_s=0.3\,\Omega$,  
 $C_{\alpha}=1.561\,\text{F\,s}^{(\alpha-1)}$, 
 $\alpha = 0.9089$; 
(b) Simulated magnitude and phase responses of the $R_s$-CPE model vs. frequency (i.e. $H_{v}(f)$ and $H_{i}(f)$). 
}
\label{fig3}
\end{center}
\end{figure}

An Eaton PowerStor supercapacitor rated 2.7\,V, 3\,F (part \#HV0820-2R7305-R) is selected for the experimental measurements and verification. Its spectral impedance plotted in terms of real part vs. imaginary part is provided in Fig.\;\ref{fig3}(a) for different dc voltage biases from 0 to 2.7\,V. The EIS measurements were carried out on a Biologic VSP\,300 station with  14.14\,mV sine perturbation ($V_{\text{rms}}=\text{10\,mV}$) on top of the dc bias, from the frequency of 1\,MHz to 10\,mHz (with 10 points per decade). Prior to the actual frequency-domain  measurements, the device was let to stabilize for a duration of 5\,min at each potential step. The results show slight variability in the impedance response as a function of the voltage bias. We fitted all low-frequency branches from 3.4\,Hz to 10\,mHz with the $R_s$-CPE model with $R_s$ preset to 0.3\,$\Omega$, and we  selected the average values  $C_{\alpha}=\text{1.561\,F\,s}^{\alpha-1}$ (standard deviation $\sigma_{C_{\alpha}}=\text{0.290\,F\,s}^{\alpha-1}$) and $\alpha = \text{0.9089}$ ($\sigma_{\alpha}=\text{0.019}$) for the subsequent calculations.
 Plots of magnitude and phase of the transfer functions $H_v(s)$ (Eq.\;\ref{eq:Hi}) and $H_i(s)$ (Eq.\;\ref{eq:Hv}) as a function of frequency from  10\,mHz to 3.4\,Hz are shown if Fig.\;\ref{fig3}(b).
 
\subsubsection{Triangular voltage excitation}

The supercapacitor device has  then been discharged into a constant resistor of 10\,$\Omega$ until its voltage reached 5\,mV, and then excited with periodic triangular voltage waveforms from 0 to 2.5\,V at different scan rates  from 0.05\,V/s to 2.0\,V/s. With those values of scan rates we expect  the device to operate within the quasi-capacitive frequency range in which its  $R_s$-CPE model parameters (i.e. $R_s =\text{0.3}\,\Omega$, $C_{\alpha}=\text{1.561\,F\,s}^{\alpha-1}$, $\alpha = \text{0.9089}$) remain valid and constant. At higher charge/discharge rates, smaller fractions of the charge can be stored and withdrawn from the supercapacitor and thus the model requires new values for the $R_s$ and CPE  parameters. Some caution should be exercised here to avoid pushing the device to a nonlinear regime when high rates are applied, which makes the modeling of the frequency response with an $R_s$-CPE circuit incorrect.   
 
\begin{figure*}[!t]
\begin{center}
\includegraphics[]{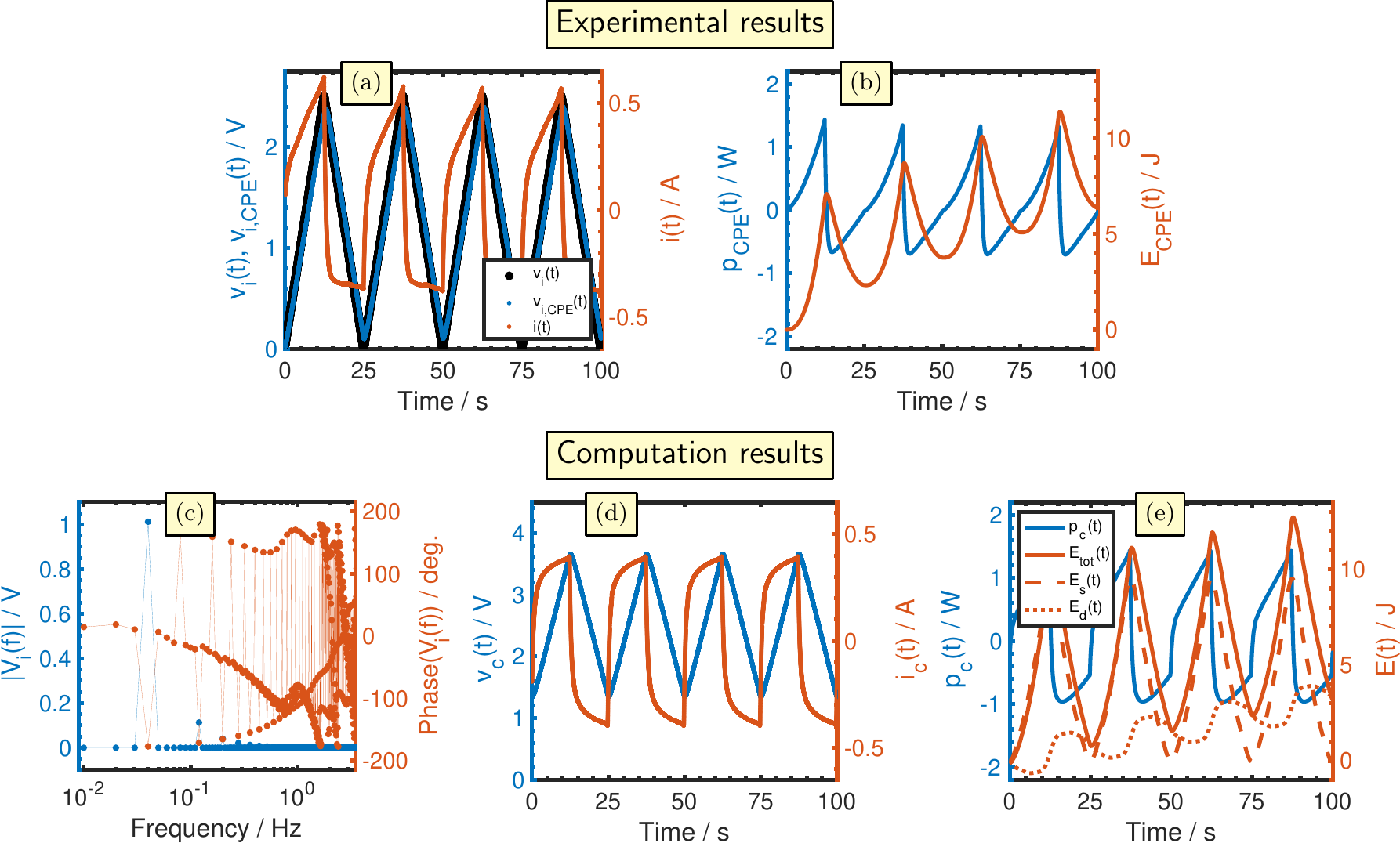}
\caption{
(a) Input triangular voltage (0 to 2.5\,V at 0.2\,V/s scan rate) and resulting current in the PowerStor supercapacitor with the voltage waveform on the CPE part only. In (b) we show the measured power and energy from which we deducted the dissipation in the series series resistance (i.e. for the CPE part only).  
(c) Plots of   magnitude and phase angle vs. frequency of the DTFT signal of $v_i(t)$ shown in (a). 
(d)  Plots of of the signals  $v_c(t)$ and $i_c(t)$ in response to $v_i(t)$ (computed via the DTFT of $v_i(t)$), and 
(e) plots of computed instantaneous  power and energy (total, stored and dissipated) on the CPE as a function of time.
}
\label{fig4}
\end{center}
\end{figure*}

Here we show the example of a 0.2\,V/s scan rate (i.e. 40\,mHz fundamental frequency) with 5\,ms sampling time, knowing that the results and conclusions obtained under other settings are comparable.  The input voltage signal $v_i(t)$ and resulting current $i(t)$ in this case are given in Fig.\;\ref{fig4}(a) for four consecutive cycles. 
The voltage on the CPE ($v_{i,\text{CPE}} = v_{i} - R_s\,i(t)$) is also given as a reference. 
  We remark that the current signal is asymmetric when looking at the charge and discharge sequences. This can be attributed to different types of materials and mechanisms of charge/discharge occurring at the (porous) anodic and cathodic sides of the device. 
In Fig.\;\ref{fig4}(b), we show the power on the CPE only (computed as $p_{\text{CPE}} = v_i(t) \,i(t) - R_s\, i^2(t) $), which is also asymmetric, as well as its time integral giving the accumulated energy on the CPE.

\begin{figure*}[!t]
\begin{center}
\includegraphics[]{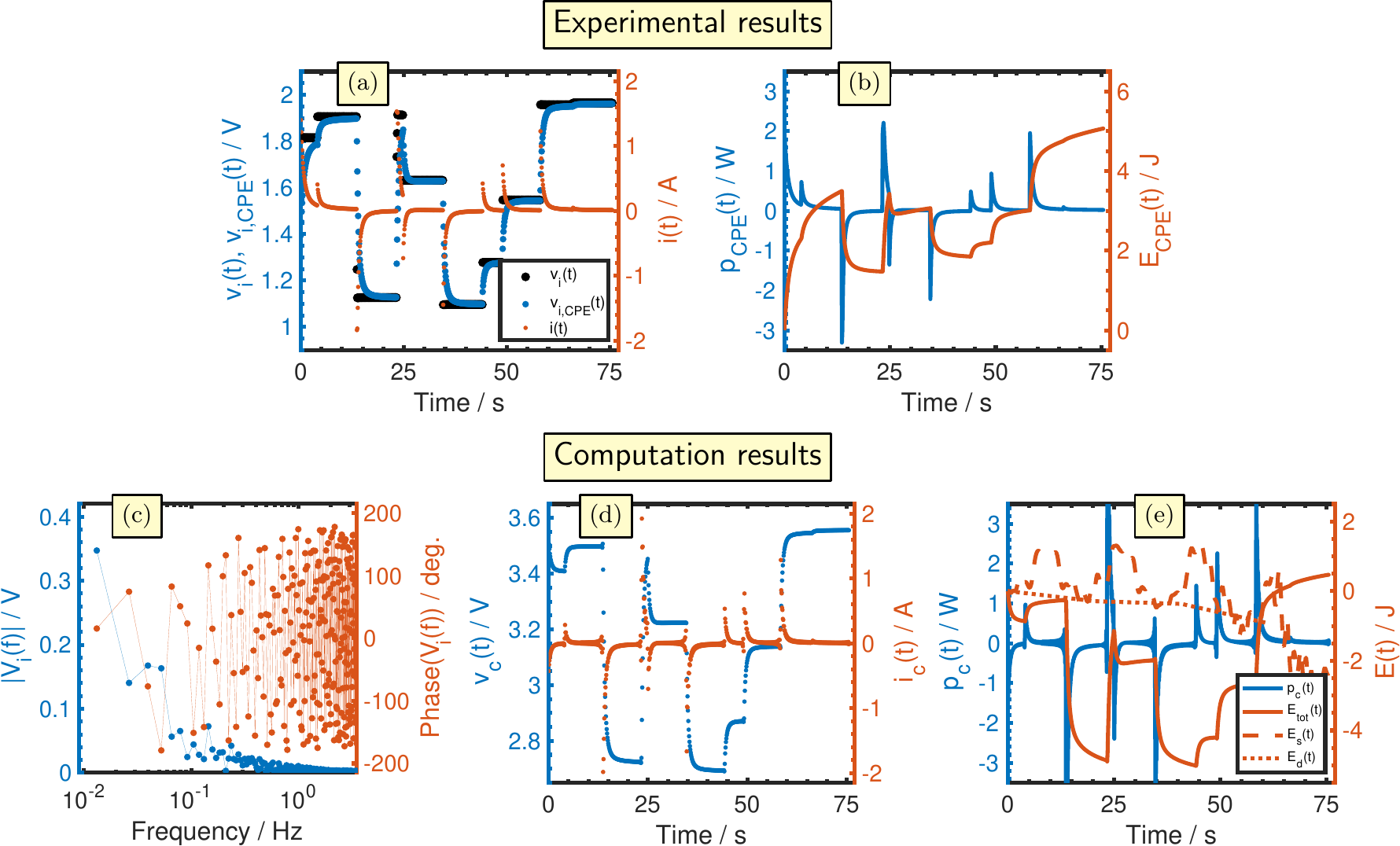} 
\caption{
(a) Input random voltage excitation of random amplitudes with random durations (voltage steps of 1.8147, 
    1.9058, 
    1.1270, 
    1.9134, 
    1.6324,
    1.0975, 
    1.2785, 
    1.5469, 
    1.9575, 
    1.9649\,V for the durations of 1.5761, 
    9.7059, 
    9.5717, 
    4.8538, 
    8.0028, 
    1.4189, 
    4.2176, 
    9.1574, 
    7.9221, 
    9.5949\,seconds, respectively) and resulting current in the PowerStor supercapacitor; we also show the result for $v_{i,\text{CPE}} = v_{i} - R_s\,i(t)$.  In (b) we show the measured power  from which we deducted the dissipation in the series series resistance (i.e. for the CPE part only), and the corresponding  energy as a function of time.  
(c) Plots of   magnitude and phase angle vs. frequency of the DTFT signal of $v_i(t)$ shown in (a). 
(d)  Plots of of the signals  $v_c(t)$ and $i_c(t)$ in response to $v_i(t)$ (computed via the DTFT of $v_i(t)$), and 
(e) plots of computed instantaneous  power $p_c(t)=v_c(t)\,i_c(t)$ and energy (total, stored and dissipated) on the CPE as a function of time.
}
\label{fig5}
\end{center}
\end{figure*}

 Now, following the computational steps of Fig.\;\ref{fig1}, we first apply the discrete-time Fourier transform (DTFT, using the fast Fourier transform  algorithm) to the input signal $v_i(t)$, which is represented in Fig.\;\ref{fig4}(c) in terms of magnitude and phase. 
Next, with Eq.\;\ref{eq:conv} we obtain the signals $v_c(t)$ and $i_c({t})$  (see Fig.\;\ref{fig4}(d)), from which the power $p_c(t)=i_{c}(t) v_{c}(t)$ is computed (see Fig.\;\ref{fig4}(e)). This, with 340 terms for computing $i_{c}(t) $ and $v_{c}(t)$, required barely 390\,ms in terms of CPU execution time. The waveforms appear to be smooth and relatively with the same trends as the measurements. However, given that we used constant values for the $R_s$-CPE model parameters for simulating both the charging and discharging sequences, it is not possible to capture the asymmetric profiles for the current signal as shown by the experiment. Again the overall trend and magnitude of the waveform is reasonably close to the experimental results, especially for the discharging steps. The  profiles of $E_{\text{tot}}$,  $E_s(t)$ and  $E_d(t)$ (computed following the same procedure presented in the previous section with five harmonics Fourier series decomposition of $p_c(t)$,  R-squared\;$=\text{0.9794}$, rmse\;$=\text{0.1217}$) are also plotted in Fig.\;\ref{fig4}(e) indicating a steady increase of all quantities with time. The   energy being stored on the CPE is more than the one being dissipated for this case of triangular voltage waveform compared to the full-wave rectified signal. The overall accumulated energy on the device is relatively comparable to that recorded experimentally (Fig.\;\ref{fig4}(b)),  but with the same remark on the asymmetry between the charge and discharge sequences.

\subsubsection{Random voltage waveform}
We repeated  the same experimental and computational procedure carried out on the PowerStor supercapacitor for the triangular voltage waveform above, but now for a non-periodic random voltage excitation. The voltage amplitudes   were derived from a uniform distribution of 1.5\,V mean value and 1.0\,V and 2.0\,V lower and upper limits, and the time durations for each voltage step were  also derived from a uniform distribution of 5.0\,s mean value and 0\,s and 10\,s lower and upper limits. The values of the ten preset voltage steps and their duration are:  1.8147, 1.9058, 1.1270, 1.9134, 1.6324, 1.0975, 1.2785, 1.5469, 1.9575, 1.9649\,V for the durations of 1.5761, 9.7059, 9.5717, 4.8538, 8.0028, 1.4189, 4.2176, 9.1574, 7.9221, 9.5949\,seconds, respectively. Data acquisition was carried out at a constant time step of 0.1\,s.
 
Plots of the experimental results in terms of input voltage and resulting current, voltage across the CPE, power and the corresponding energy on the CPE  as a function of time are all provided in Figs.\;\ref{fig5}(a) and\;\ref{fig5}(b). Plots of the results obtained from the computational procedure  are summarized in the second row of Fig.\;\ref{fig5}. 
The computed current waveform shown in Fig.\;\ref{fig5}(d) is in very good agreement with the experimentally-measured current shown in  Fig.\;\ref{fig5}(a).  
 However, when comparing the waveforms for the voltage on the CPE ($v_{i,\text{CPE}} = v_{i} - R_s\,i(t)$) obtained from the measurements (Fig.\;\ref{fig5}(a)) and the computed voltage $v_c(t)$ (Fig.\;\ref{fig5}(d)) we realize that   there is a clear dc upward shift for $v_c(t)$ that can be attributed  to errors in the computation of the low-frequency DTFT magnitude and phase spectra  of the (non-periodic) input voltage. The transitions from one step  to another are, on the other hand, properly captured by the simulations.   
As a result we see in Fig.\;\ref{fig5}(e) that the estimated power for the CPE part of the device shows higher peaks at these instances when compared to the actual measurements (Fig.\;\ref{fig5}(b)), but the overall distribution and timing of each sequence are as expected.  Note that the  computational steps up to the calculation of the power $p_c(t)$ took 50\,ms for this case (381 summation terms for each of $v_c(t)$ and $i_c(t)$). 
 The same remarks for the power waveform can be said for $E_{\text{tot}}$ (see Fig.\;\ref{fig5}(e)) when compared to the experimental $E_{\text{CPE}}$ from  Fig.\;\ref{fig5}(b). 
 The evolution of estimates for  $E_s(t)$ and  $E_d(t)$ as a function of time for this case are also computed (with Eqs.\;\ref{eqEs} and\;\ref{eqEd}) and plotted in  Fig.\;\ref{fig5}(e). Here we have to mention that it was difficult to obtain satisfactory Fourier series approximation of the highly-irregular power waveform (R-squared was as low as 0.5016, and the rmse was 0.7716 using 200 harmonics) making the results to be viewed just as an indicative figure. 

\section{Conclusion}

We described a systematic procedure to estimate the time-domain power and stored and dissipated energies in supercapacitors in a response to arbitrary voltage excitation with only prior knowledge of their measured impedance parameters. The procedure is based on the convolution theorem for frequency-time transformation, and the approximation of the excitation signal with Fourier series, which makes it versatile and readily applicable to any impedance function and input signal. 
We examined and verified the response of a fractional-order $R_s$-CPE circuit approximated by a fifth-order distributed $RC$ equivalent circuit. We also tested the procedure on a commercial supercapacitor behaving as an $R_s$-CPE circuit using both a periodic triangular voltage excitation and a non-periodic random voltage excitation. The results are very promising for implementing and using such method in real-world applications for estimating supercapacitor responses in terms of current, voltage and power capabilities.

%
%
%

\bibliographystyle{IEEEtran} 



 
\vspace{11pt}

\vspace{11pt}

\vfill

\end{document}